# *In situ* photoluminescence and Raman study of nanoscale morphological changes in organic photovoltaics during solvent vapor annealing

Giovanni Fanchini*, Steve Miller* and Manish Chhowalla
*These two authors contributed equally to this work.

Improvement of the photovoltaic efficiency (from 1.2% to 3%) via exposure of organic poly(3-hexylthiophene) (P3HT) + phenyl-$C_{61}$-butyric acid methyl ester (PCBM) devices to solvent vapor at room temperature is reported. *In situ* photoluminescence (PL) and Raman spectroscopies, in conjunction with optical absorption data, have been used to provide insight into the nanoscale morphological changes occurring during solvent vapor annealing. We found that in P3HT:PCBM, suppression of PL and the decrease in line-width of the 1442 cm$^{-1}$ P3HT Raman peak are accompanied by strong modifications in the optical absorption spectra during solvent vapor annealing, in contrast to measurements on P3HT only films. We attribute these observations to de-mixing of PCBM and subsequent stacking of P3HT in coplanar conjugated segments, similar to what is observed during thermal annealing.

**Introduction**

Organic photovoltaics are promising alternatives to their inorganic counterparts [1] because they can be prepared simply from solution on cheap substrates which in principle should to compensate for their lower efficiency. The most successful organic photovoltaic materials system exploits the properties of the soluble fullerene derivative, phenyl-$C_{61}$-butyric acid methyl ester (PCBM), as an electron acceptor and a conducting polymer, poly(3-hexylthiophene) (P3HT), as an electron donor [1-14]. Efficiencies of these bulk heterojunction solar cells have been reported to be around 5% [13,14] when devices are prepared on rigid substrates in inert environments. There is great experimental interest in P3HT:PCBM photovoltaic devices because theoretical studies have predicted efficiencies as high as 11% [15].

Thermal annealing of the P3HT – PCBM bulk heterojuction devices is a critical step for improving the efficiency [13]. It has been demonstrated (through post experimental analysis) that thermal annealing strongly affects the morphology of the nanocomposite layer via de-mixing of PCBM [16] and stacking of P3HT in coplanar conjugated segments [17]. In thermal annealing, relatively high temperatures from 120 $^{o}$C to 150$^{o}$C are often used [1,2,13], which requires care for devices fabricated on cheap flexible plastic substrates. Recently Dickey et al [18] reported an improvement in the performance of organic 3-ethyl-silylethynyl anthradithiophene (TES ADT) transistors via room temperature solvent vapor annealing (that is, exposing the deposited organic layers to solvent vapor). The beneficial effect of better structural order in TES ADT through solvent vapor annealing suggests that it could also improve the performance of P3HT – PCBM solar cells.

In this report, the improvement in the photovoltaic efficiency and structural changes occurring during solvent vapor annealing of flexible P3HT – PCBM solar cells in chloroform (CHCl$_3$) vapor at room temperature are described. Specifically, the efficiencies of our devices fabricated and tested in air increases from 1.2% (for as fabricated devices) to 3% after solvent vapor annealing which is comparable to values obtained after thermal annealing. Furthermore, the maximum efficiency saturates after approximately 30 s and longer annealing times of up to several minutes does not lead to significant improvement. In order to gain insight into the mechanisms responsible for the improved efficiency within the first 30 s, we have investigated structural changes in the nanocomposite layer during solvent vapor annealing. Although CHCl$_3$ vapor is known to improve the electroluminescence in P3HT [19], to the best of our knowledge, we are not are not aware of any other studies in which solvent vapor annealing has been demonstrated to be as effective as thermal annealing in improving the efficiency of organic photovoltaics and that *in – situ* monitoring of structural changes in the P3HT – PCBM layer during annealing has been reported.

**Experimental Details**

The organic solar cells were prepared by depositing a mixture of P3HT (regioregularity 98.5%, Riekie Metals Inc.) and PCBM on to indium-tin oxide (ITO) layer (<20 Ω·□ resistance, >85% transmittance) on Poly(ethylene terephthalate) substrates covered with a 10-nm thick poly(3,4-ethylene dioxy-thiophene):poly(styrenesulfonate) (PEDOT:PSS). Specifically, a 10 mg/mL solution at 1:1 weight ratio of P3HT to PCBM in CHCl$_3$ was spin-coated at 1000 rpm, resulting in a 100 – 120 nm thick active layer. Ga-In eutectic was used for the top contact, as described elsewhere [20]. The photovoltaic characteristics were evaluated under a Newport AM1.5 solar simulator.

The *in situ* morphological and electronic changes in the P3HT – PCBM layers were observed by monitoring the changes in the Raman and PL features using a Renishaw InVia spectrometer operating at 1.57 eV excitation. Each spectrum required 3 s to record and care was taken to avoid laser soaking of the sample. The experimental configuration used to perform solvent vapor annealing while simultaneously measuring the PL and Raman features is shown in Figure 1a. As special apparatus was fabricated in which the substrate placed upside down was the top cap of a Petri dish which was filled with chloroform. This allowed the laser to pass through the PET and ITO layers from the top to the P3HT:PCBM layer while being exposed to the solvent vapor. The choice of a low excitation energy ($E_i$ = 1.57 eV) strongly favors the observation of the Raman and PL arising from P3HT while suppressing the PL from PCBM which has a much larger optical gap. Furthermore, using such a low excitation energy in a spectral region where P3HT is transparent allows the determination of the shape and intensity of the PL peak without any corrective models accounting for the reabsorption of luminescence by the sample itself. Such models are unreliable in the presence of strong changes in the optical absorption edges observed during solvent vapor annealing (see below).

PL in conjugated thiophenes is well known [21-23] to arise from radiative recombination of polaron-exciton pairs into Franck-Condon (FC) states, as sketched in Figure 1b. The shift of the PL peak with respect to the excitation energy provides information about the polaron-exciton binding energy, (2U, as indicated in Figure 1b) [23]. Therefore, PL spectroscopy can be a powerful tool for the investigation of the nature, the binding energy and the degree of confinement of the electronic states in photoexcited P3HT. It should be noted that for some *in situ* Raman and PL measurements, the PEDOT:PSS layer was omitted in order to avoid masking of P3HT features which overlap with those of PEDOT. Additional films of pure P3HT were also deposited and analyzed for comparison. Tapping mode atomic force microscopy (AFM) (Digital Instruments Nanoscope) and optoelectronic measurements (using a Perkin-Elmer Lambda 20 transmittance spectrophotometer for the solutions and a Jobin-Yvon UVISEL spectral ellipsometer coupled with a multilayer model for the solid films) were also performed.

**Results and Discussion**

The current density versus voltage curves under AM1.5 illumination for as fabricated and solvent vapor annealed photovoltaic devices are shown in Figure 2. It can be observed that the open circuit voltage decreases slightly upon solvent vapor annealing while the short circuit current increases by two fold (from ~ 3 to 6 mA/cm$^2$). The efficiency as a function of the solvent vapor annealing time is plotted in the inset of Figure 2. It can be seen that after approximately 30 s, the efficiency saturates and remains at a maximum value of approximately 3%. Efficiency of thermally annealed samples is also plotted for comparison. In contrast to other parameters such as the short circuit current and fill factor which increase then saturate, the open circuit voltage circuit was found to decrease from 0.65 V to 0.59 V after the initial 30 s of solvent and saturate, as also shown in the other inset of Figure 2. We further discuss the decrease in open circuit voltage below.

In order to establish the mechanism for the enhancement in efficiency by solvent vapor annealing, we have performed additional characterization of the devices. Qualitatively, we observed that the color of P3HT films spin-coated in the absence of PCBM is purple and does not change upon solvent vapor annealing. In contrast, the P3HT-PCBM films are orange immediately after spin coating but become purple during the first 30 s of solvent vapor annealing. These effects can be quantitatively measured using optical absorption spectroscopy. The optical absorption spectra of as deposited thin films and their respective starting solutions are compared in Figure 3a. In order to compare the samples, we have adopted the $E_{04}$ parameter to quantitatively identify the optical absorption edges [24]. It can be observed that the absorption edges of the P3HT:PCBM solution (550 nm) and thin films (570 nm or 2.17 eV) are similar. In contrast, the optical absorption edge of the P3HT films strongly red-shift to 635 nm (1.95 eV) which is much lower than the $E_{04}$ of the starting solution (560 nm or 2.21 eV). The effect of solvent vapor annealing on the absorption spectra of the thin films is compared in Figure 3b. The optical absorption spectra of pure P3HT, which are red-shifted during solidification (Figure 3a) are insensitive to solvent vapor annealing. In contrast, the $E_{04}$ of P3HT:PCBM films, which weakly changed upon deposition, dramatically red-shifts upon solvent vapor annealing and approaches the value of pure P3HT (615 nm or 2.02 eV). It can also be observed from Figure 3b that the absorption features related to PCBM do not change upon solvent vapor annealing and are present only below 400 nm (i.e. > 3 eV), in agreement with literature [25]. Thus, the changes in the observed optical absorption edges, during solidification and solvent vapor annealing are manly due to structural modifications in the P3HT, in agreement with observations of Chirvase et. al [11]. Such a change, although prevented during the deposition of P3HT:PCBM films, is favored during solvent vapor annealing and provides beneficial effects on the photovoltaic device efficiencies.

This interpretation is consistent with a similar explanation for structural modifications of P3HT-PCBM during thermal annealing [14,16]. In particular, Yang et al [16] showed by means of transmission electron micrographs (TEM) that thermal annealing leads to de-mixing of P3HT and PCBM by forming PCBM-rich regions, allowing the reorganization of the residual P3HT phase in stacked coplanar conjugated segments. The reorganization is prevented in the as deposited P3HT:PCBM films because of the rapid solidification and intercalation of PCBM. In solvent vapor annealed P3HT:PCBM however, this ordering phenomenon can occur via the diffusion of $CHCl_3$ vapor which softens the organic nanocomposite, allowing the migration of PCBM and the subsequent reorganization of P3HT. Indeed, the solvent vapor annealing may be comparable to controlled evaporation of the solvent after deposition which has also been demonstrated to improve the solar cell efficiency, presumably by allowing for ordering to occur [14]. We suspect that the observed changes in the optical absorption edge may also be due to stacking and subsequent π-conjugation of the P3HT segments.

The physical changes in the morphology of the P3HT-PCBM films before and after solvent vapor annealing observed by AFM are shown in Figure 3c and 3d, respectively. The as deposited thin films appear featureless while clusters of 10-20 nm are prominent on the surface of the solvent annealed films. The AFM image in Figure 2d is comparable to the images reported in the literature [1,5] which show dramatically improved efficiencies. It should be noted that no change in morphology by AFM upon solvent vapor annealing was observed in pure P3HT films.

The PL and Raman spectra of P3HT and P3HT-PCBM films before and after solvent vapor annealing are shown in Figure 4a. It can be seen from Figure 4a that the polaron-exciton binding energy in our samples remains at 2U ~ 0.2 eV, irrespective of the annealing time and the presence of PCBM. In P3HT, the polaron-exciton confinement parameter ($\gamma$) can be extracted from the $E_{04}$ energy in the absorption spectrum and the correlation energy using the formula [23,27]:

$$\gamma = (1-2U/E_{04}) \cdot \sin^{-1}(1-2U/E_{04}) \cdot [1 - (1-2U/E_{04})^2]^{-1/2} \qquad (1)$$

Substituting our $E_{04}$ values from measurements in Figures 3a and 3b in equation 1, $\gamma$ is found to remain constant (decreases from 0.44 to 0.43). This indicates that the exciton confinement in P3HT is not affected by the annealing time or by the presence of PCBM, similar to other reports in the literature [21,28].

The evolution of the PL peak intensity as a function of the solvent vapor annealing time is plotted in Figure 4b. Unlike the degree of confinement which was not affected by the presence of PCBM or by solvent vapor annealing, the PL intensity is strongly quenched by solvent vapor annealing in P3HT:PCBM films. Furthermore, the decrease in PL in the P3HT:PCBM films mostly occurs in the first 30 s of solvent vapor annealing, correlating to the improvement in the device efficiency. Note that although the PL intensity also decreases in P3HT films but the effect is less dramatic than in the P3HT – PCBM material.

Returning to Figure 4a, a number of Raman peaks superimposed on the broad PL spectra can also be observed. Raman spectra recorded from bare substrates allow us to assign the peaks marked with stars (*, **) to PET and ITO, respectively. The 715 $cm^{-1}$, 1380 $cm^{-1}$ and 1440 $cm^{-1}$ Raman peaks indicated by squares are assigned to the P3HT C-S-C ring deformation, C-C skeletal stretching and C=C ring stretching, respectively [29]. No Raman features attributable to PCBM, such as the $A_{1g}$ 1469 cm$^{-1}$ mode of fullerenes could be resolved. In order to extract additional information from the Raman spectra, we compare the 1350-1500 cm$^{-1}$ region of P3HT and P3HT – PCBM devices before and after solvent vapor annealing in Figure 4c. It can be seen from Figure 4c that while the Raman shift of the C=C peak is almost constant, interesting effects on the full width at half maximum (FWHM) of the peak can be observed. In pure P3HT, the FWHM decreases insignificantly (from 24 cm$^{-1}$ to 20 cm$^{-1}$) with solvent vapor annealing, as shown in Figure 4d. This minor change in the FWHM is consistent with the negligible changes observed in the optical absorption spectra, sample morphology and the insignificant decrease in PL intensity. In contrast, in P3HT – PCBM films, the C=C signal before solvent vapor annealing is much broader (about 32 cm$^{-1}$) and after solvent vapor annealing it narrows to 22 cm$^{-1}$, close to the 20 cm$^{-1}$ FWHM of pure P3HT films. Once again, the relatively more dramatic decrease in the FWHM occurs within the first 30 s of solvent vapor annealing time. Furthermore, since the Gaussian widths of the Raman lines in heterogeneous systems are a measure of their disorder [30], we can infer that solvent vapor annealing strongly reduces the disorder in P3HT – PCBM, consistent with stacking of conjugated P3HT segments.

Based on the above data, it is clear that solvent vapor annealing of P3HT – PCBM leads to substantial ordering in the polymer so that the optoelectronic, morphological, PL and Raman characteristics are comparable to highly ordered P3HT. This is due to the strong reduction in disorder via segregation of PCBM into PCBM-rich regions and stacking of P3HT chains. These structural changes have as a beneficial effect on the solar cell performance (see Figure 2) which can be explained by the enhanced transport of the carriers through ordering of the P3HT – PCBM matrix [16]. Furthermore, as demonstrated by Bässler and co-workers [31] and Koehler and co-workers [32], the morphology of the interface, which will also be affected by solvent vapor annealing, is also important to the process of charge carrier dissociation.

The improvement in the solar cell device efficiency may also arise from an additional mechanism occurring during solvent vapor annealing. Based on our measurements, we have developed the energy diagram shown in Figure 5 for the photovoltaic devices. In the as deposited P3HT:PCBM films (dotted line in the P3HT energy levels in Figure 5), the presence of PCBM prevents the stacking of P3HT so that the optical gap and excitonic levels are similar to those of P3HT in solution [Chirvase, Natotechnology]. Specifically, assuming a work function of 4.3 eV for P3HT, the lowest unoccupied molecular orbital (LUMO) equilibrium level for unannealed P3HT is at $E_1$ [$S_{1,(eq)}$] ~ 3.3 eV compared to the acceptor level of PCBM at ~ 3.75 eV [33,34]. Assuming these energy levels, the probability ($p \sim \exp[-(E_2 - E_1)/k_BT]$) for an electron in the LUMO of P3HT to diffuse into the PCBM at room temperature ($k_BT \approx 0.025$ eV) is rather low. Thus, the electron-hole pair tends to radiatively recombine in P3HT generating the strong PL observed in Figure 4a. The exciton delocalization in our devices is not substantially improved, as indicated by the negligible change in the correation energy (U) and localization parameter ($\gamma$), after solvent vapor annealing. However, the decrease in the optical gap (from absorption spectroscopy) indicates that the LUMO of P3HT after solvent vapor annealing is closer to the acceptor state of the PCBM, as shown in Figure 5. Therefore, although the diffusion probability is still quite low, it is now exponentially increased, making the transition indicated by the dark arrows in Figure 5 more likely. Furthermore, since the open circuit voltage in P3HT:PCBM devices is more closely related to the differences in bulk heterojunction energies rather than the difference between the work functions of the two electrodes [35], it is expected to decrease with solvent vapor annealing based on Figure 5. This is clearly observed in the inset of Figure 2 in which the open circuit voltage is seen to rapidly decrease within the first 30 s of solvent vapor annealing and then saturate.

**Conclusions**

In conclusion, we have demonstrated that solvent vapor annealing of P3HT:PCBM at room temperature leads to a marked increase in the photovoltaic efficiency. We have correlated the enhanced efficiency to the ordering of P3HT, presumably due to demixing of PCBM as observed by TEM [7]. The more ordered structure provides better transport for holes through the more crystalline P3HT and electrons via hopping through the PCBM, comparable to what occurs during thermal annealing. The dissociation of excitons and ordering of the P3HT have been observed using *in – situ* photoluminescence and Raman spectroscopies. We discovered that the morphological changes which correlate to the improvement in the efficiencies occur within the first 30 s of the solvent vapor annealing. Exposure to the solvent vapor for longer periods does not change the structure or the device performance. Therefore, solvent vapor annealing might be an alternative to thermal annealing for improving the P3HT-PCBM photovoltaic performance, especially for temperature sensitive substrates.

Acknowledgements

We would like thank Prof. Adrian Mann for allowing use of the Raman instrument. We also acknowledge financial support from National Science Foundation CAREER Award (ECS 0543867).

References:

1. S-S.Sun, N.S.Sariciftci (eds.) Organic Photovoltaics, Taylor & Francis, London, 2005.

2. C.J. Brabec, V. Dyakonov, J. Parisi, N.S. Sariciftci (Eds.), Organic Photovoltaics: Concepts and Realization, Springer Verlag, Heidelberg, 2003.

3. N.S. Sariciftci, L. Smilowitz, A.J. Heeger, F. Wudl, Science 258 (1992) 1474.

4. G. Yu, J. Gao, J.C. Hummelen, F. Wudl, A.J. Heeger, Science, 270 (1995) 1789.

5. S.E. Shaheen, C.J. Brabec, N.S. Sariciftci, F. Padinger, T. Fromherz, J.C. Hummelen, Appl. Phys. Lett. 78 (2001) 841.

6. C.J. Brabec, J.C. Hummelen, N.S. Sariciftci, Adv. Funct. Mater. 11 (2001) 15.

7. H. Sirringhaus, N. Tessler, R.H. Friend, Science 280 (1998) 1741.

8. N. Camaioni, G. Ridolfi, G.C. Miceli, G. Possamai, M. Maggini, Adv. Mater. 14 (2002) 1735.

9. D. Chirvase, Z. Chiguvare, M. Knipper J. Parisi, V. Dyakonov, J.C. Hummelen, J. Appl. Phys. 93 (2002) 3376

10. F. Padinger, R.S. Rittberger, N.S. Sariciftci, Adv. Funct. Mater. 13 (2003) 85.

11. D. Chirvase, J. Parisi, J.C. Hummelen, V. Dyakonov, Nanotechnology, 15 (2004) 1317.

12. I. Riedel, V. Dyakonov, Phys. Stat. Sol. (a) 201 (2004) 1332.

13. M. Reyes-Reyes, K. Kim, D.L. Carroll Appl. Phys. Lett. 87 (2005) 083506.

14. G.Li, V.Shrotriya, J.Huang, Y.Yao, T.Moriarty, K.Emery, Y.Yang, Nature Materials 4 (2005) 864

15. L.J.A. Koster, V.D. Mihailetchi, P.W.M. Blom, Appl. Phys. Lett. 88,

093511 (2006).

16. X. Yang, J. Loos, S.C. Veenstra, W.J.H. Verhees, M.M. Wienk, J.M. Kroon, M.A.J. Michels, R.A.J. Janssen, Nano Lett. 5 (2005) 579

17. N.Camaioni, G. Ridolfi, G. Casalbore-Miceli, G. Possamai, M. Maggini, Adv. Mater. 14 (2002) 1735.

18. K.C. Dickey, J.E. Anthony, Y.-L. Loo, Adv. Mater. 18 (2006) 1621.

19. M. Berggren, G. Gustafsson, O. Inganas, M.R. Andersson, O. Wennerstrom, T. Hjertberg, Appl. Phys. Lett. 65 (1994) 1489

20. A. Du Pasquier, S. Miller, M. Chhowalla, Sol Energy Mats Sol Cells 90 (2006) 1828.

21. Y.H. Kim, D. Spiegel, S. Hotta, A.J. Heeger, Phys. Rev. B 38 (1988) 5490.

22. B.C. Hess, G.S. Kanner, Z. Vardeny, Phys. Rev. B 47 (1993) 1407.

23. B. Xu, S. Holdcroft, Macromolecules 26 (1993) 4457.

24. The $E_{04}$, a quantity often used to define a gap in semiconductors, is the photon energy or wavelength at which the absorption coefficient reaches $10^4$ cm$^{-1}$. It is a thickness-independent quantity approximately representing the maximum photon energy at which a micrometric film is transparent.

25. V Shrotriya, J Ouyang, R J Tseng, G Li and Y Yang, Chem. Phys. Lett. 411 (2005) 138.

26. J. Ruhe, N.F. Colaneri, D.D.C. Bradley, R.H. Friend, G.J. Wegner, J. Phys. : Cond. Matter 2 (1990) 5465.

27. O.A. Patil, A.J. Heeger, F. Wudl, Chem. Rev. 88 (1988) 183.

28. Z. Bao, A. Dodabalapur, A. Lovinger, Appl. Phys. Lett. 69 (1996) 4108.

29. M. Baibirac, M. Lapkowski, A. Pron, S. Lefrant, I. Baltog, J. Raman Spectrosc. 29 (1998) 825.

30. M.V.Klein in *Light Scattering in Solids* Vol.8 (M.Cardona ed) Springer Berlin, p148.

31. V. I. Arkhipov, P. Heremans, and H. Bässler, Appl. Phys. Lett. 82, 4605 (2003).

32. M. Koehler, M. C. Santos, M. G. E. da Luz, J. Appl. Phys. 99, 053702 (2006).

33. D. Chirvase, Z. Chiguvare, M. Knipper, J. Parisi, V. Dyakonov, and J. C. Hummelen, Synth. Met. 1, 10348 (2003).

34. R. Valaski, L. M. Moreira, L. Micaroni, and I. A. Hummelgen, J. Appl. Phys. 92, 2035 (2002).

35. Christoph J. Brabec, Antonio Cravino, Dieter Meissner, N. Serdar Sariciftci, Thomas Fromherz, Minze T. Rispens, Luis Sanchez, and Jan C. Hummelen, Advanced Functional Materials 11, 374 (2001).

Figure Captions:

Figure 1: (a) Schematic of the experimental configuration used to perform solvent vapor annealing while simultaneously measuring the PL and Raman features. (b) Schematic of radiative recombination mechanism of polaron-exciton pairs into Franck-Condon (FC) states (adapted from Ref. 23).

Figure 2: Current density versus the voltage plot for P3HT – PCBM photovoltaic devices before (open circles) and after 30 s of solvent vapor annealing (closed red circles). The solar cell efficiency and the open circuit voltage versus the solvent vapor annealing time is plotted in the upper left and right insets, respectively.

Figure 3: (a) Comparison of absorption spectra of P3HT and P3HT – PCBM in solution and as deposited thin solidified films. (b) Comparison of absorption spectra of as deposited thin films and solvent annealed thin films of P3HT and P3HT – PCBM. The dotted horizontal line in both graphs indicates the $E_{04}$ energy used to compare the absorption coefficients among the various samples. (c) AFM image of as deposited and (d) solvent vapor annealed P3HT – PCBM thin films.

Figure 4: (a) PL and Raman spectroscopy results of pure P3HT and P3HT:PCBM before and after solvent vapor annealing. (b) Summary of the PL intensity versus the annealing time for pure P3HT and P3HT:PCBM. (c) The 1350-1500 cm$^{-1}$ Raman region of P3HT and P3HT – PCBM devices before and after solvent vapor annealing. (d) Summary of the FWHM of the 1440 cm$^{-1}$ Raman C=C peak versus the annealing time for pure P3HT and

P3HT:PCBM.

Figure 5: Energy level diagram based on our measurements before and after solvent vapor annealing. The P3HT LUMO level shifts downward as indicated by the red arrows, increasing the possibility of electron transfer into PCBM and hole transfer into ITO, as indicated by the black arrows.

Figure 1: Fanchini et al.

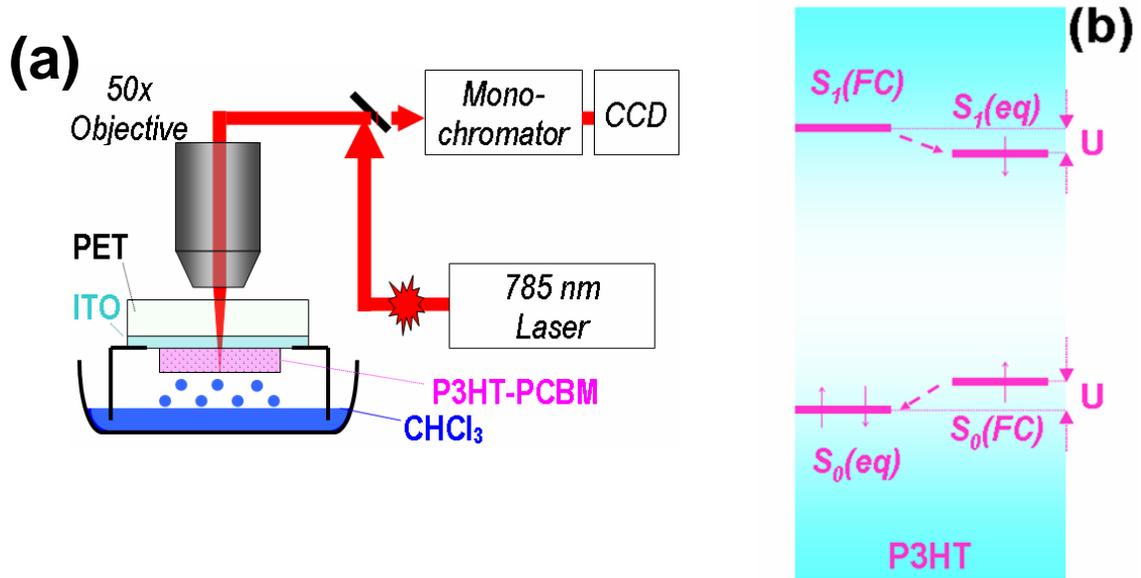

Figure 2: Fanchini et al.

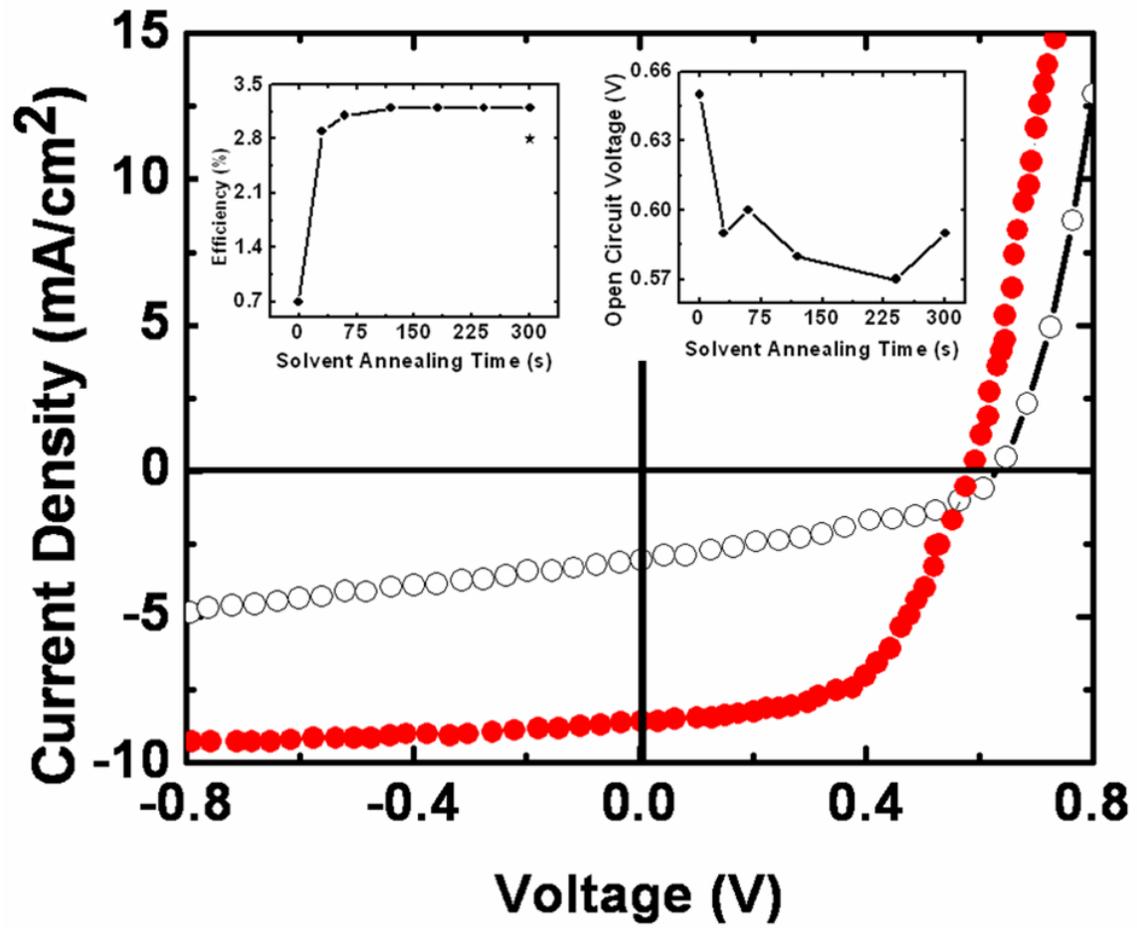

Figure 3: Fanchini et al.

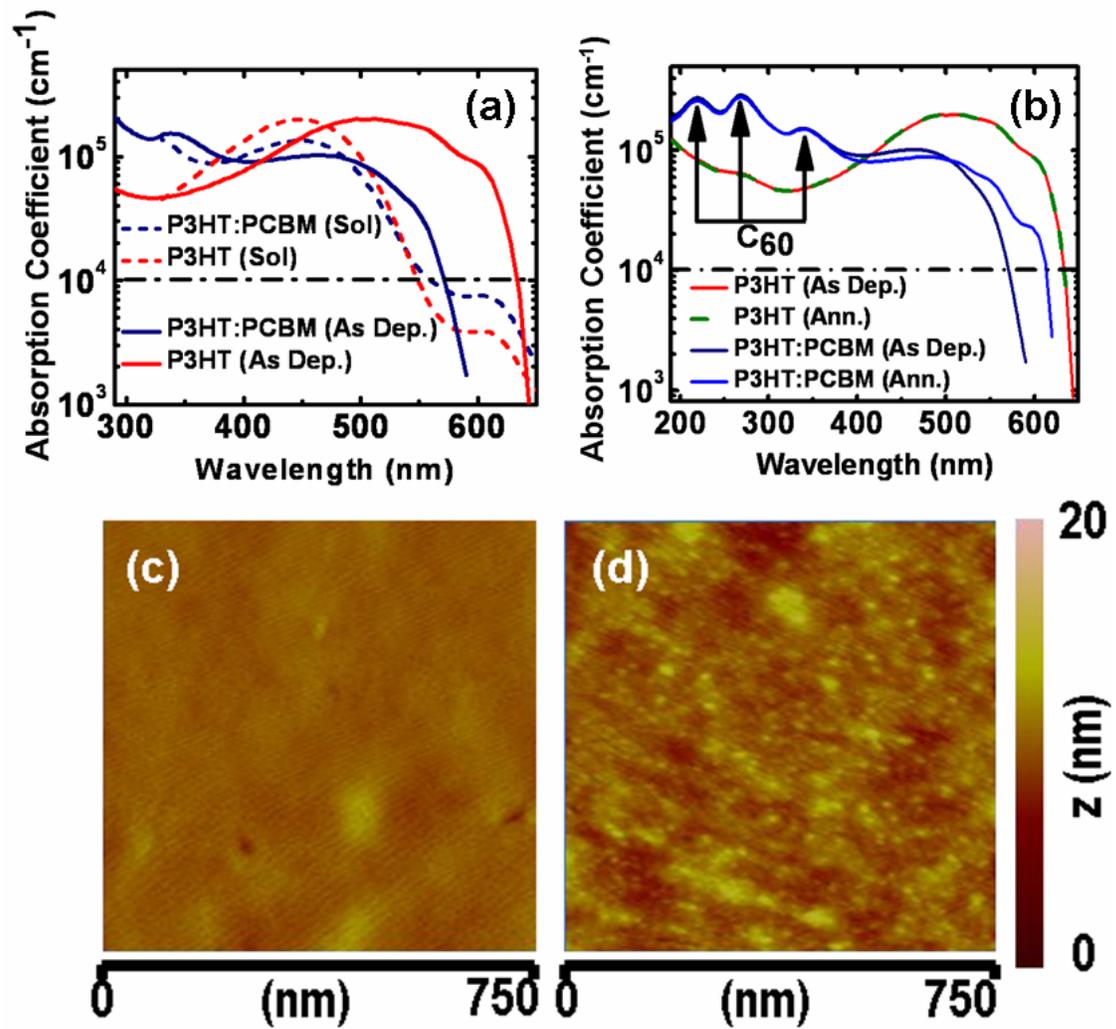

Figure 4: Fanchini et al.

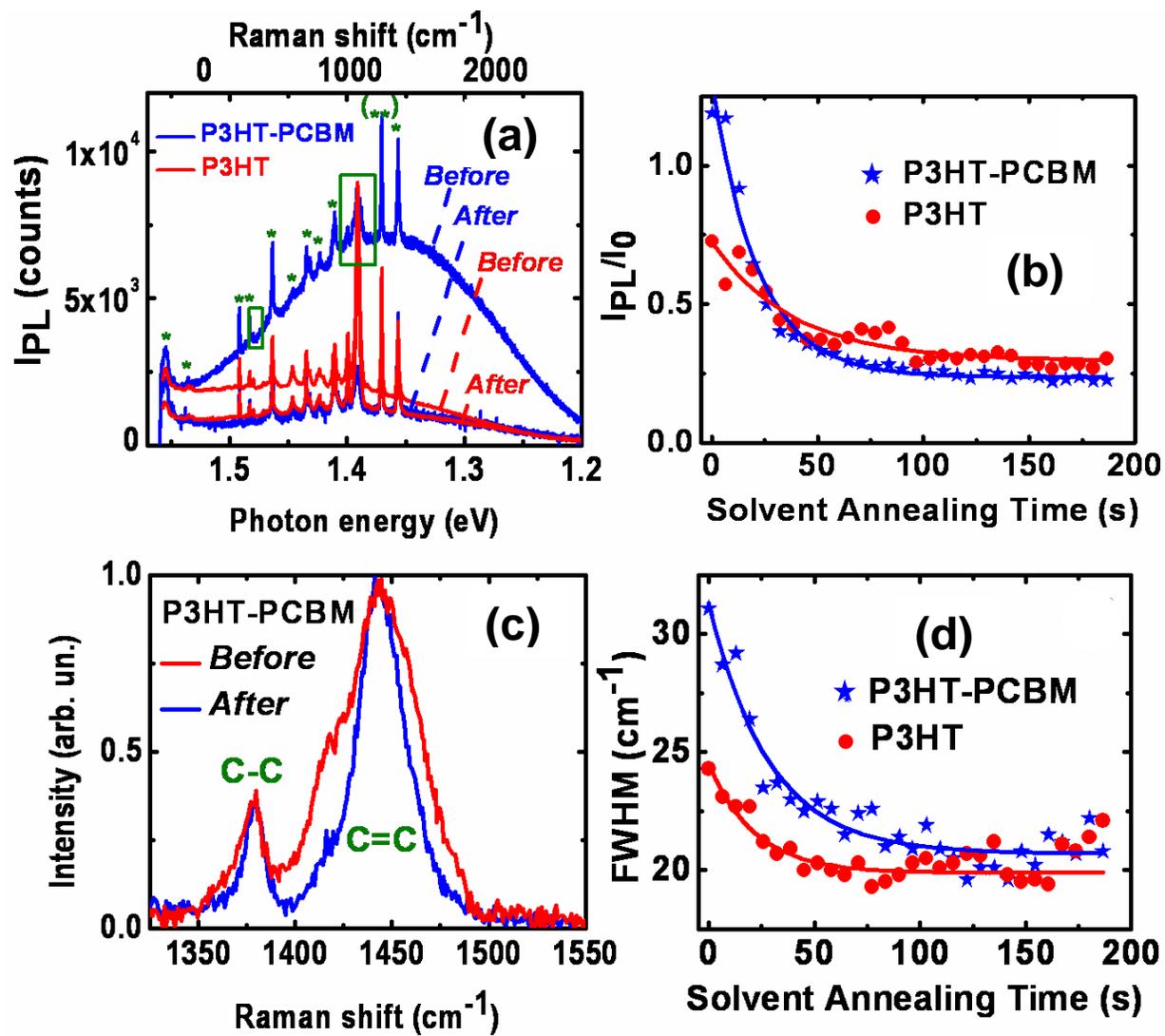

Figure 5: Fanchini et al.

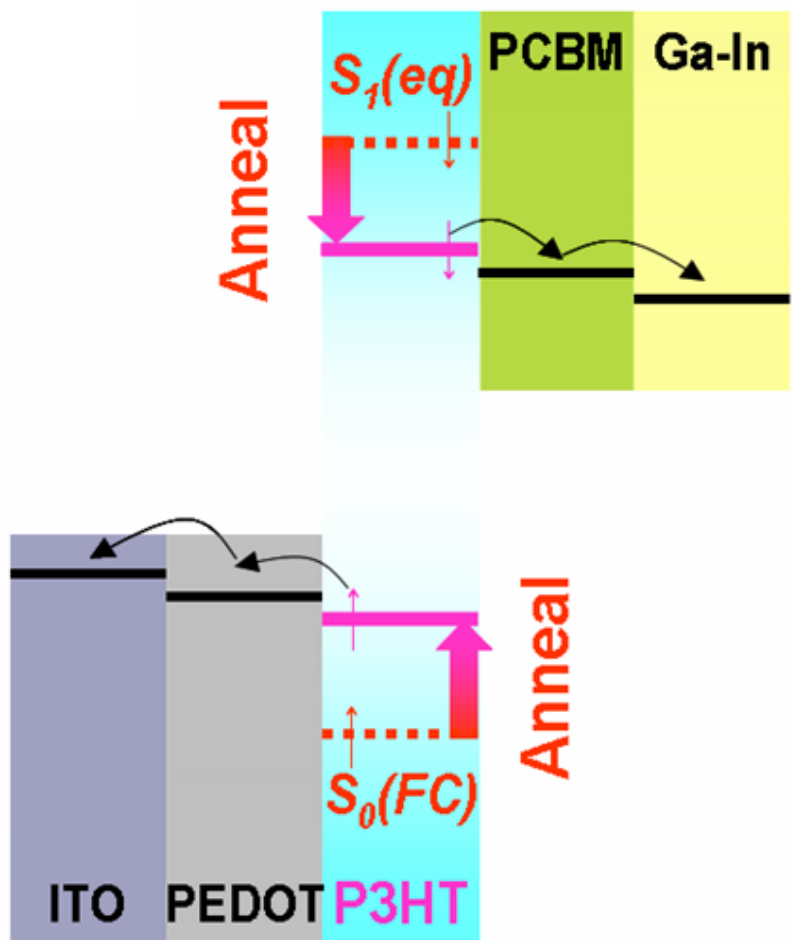